\documentclass[journal]{IEEEtran}
\usepackage{color}
\usepackage{dcolumn}
\newcolumntype{d}[1]{D{.}{.}{#1}}

\ifCLASSINFOpdf
   \usepackage[pdftex]{graphicx}
\else
\fi
\usepackage{amsmath}
\usepackage{bm}
\ifCLASSOPTIONcompsoc
  \usepackage[caption=false,font=normalsize,labelfont=sf,textfont=sf]{subfig}
\else
  \usepackage[caption=false,font=footnotesize]{subfig}
\fi
\begin{document}
\title{Tracing carbon dioxide emissions in the European electricity markets}%

\author{\IEEEauthorblockN{
Mirko Sch{\"a}fer\IEEEauthorrefmark{1},
Bo Tranberg\IEEEauthorrefmark{2},
Dave Jones\IEEEauthorrefmark{3} and
Anke Weidlich\IEEEauthorrefmark{1}}\\
\IEEEauthorblockA{\IEEEauthorrefmark{1}Department of Sustainable Systems Engineering (INATECH), University of Freiburg, 79110 Freiburg, Germany\\
Email: mirko.schaefer@inatech.uni-freiburg.de}\\
\IEEEauthorblockA{\IEEEauthorrefmark{2}Ento Labs ApS, 8000 Aarhus, Denmark}\\
\IEEEauthorblockA{\IEEEauthorrefmark{3}Ember, London, Great Britain}%

\thanks{
{\copyright} 2020 IEEE. Personal use of this material is permitted. Permission from IEEE must be obtained for all other uses, in any current or future media, including reprinting/republishing this material for advertising or promotional purposes, creating new collective works, for resale or redistribution to servers or lists, or reuse of any copyrighted component of this work in other works.}
}
\IEEEoverridecommandlockouts
%\IEEEpubid{978-1-7281-1257-2/19/\$31.00 {\copyright}2019 IEEE }

%\markboth{16th International Conference on the European Energy Market}%
%{Shell \MakeLowercase{\textit{et al.}}: Principal Cross-Border Flow Patterns in the European Electricity Market}

\maketitle

\begin{abstract}
Consumption-based carbon emission measures aim to account for emissions associated with power transmission from distant regions, as opposed to measures which only consider local power generation. Outlining key differences between two different methodological variants of this approach, we report results on consumption-based emission intensities of power generation for European countries from 2016 to 2019. We find that in particular for well connected smaller countries, the consideration of imports has a significant impact on the attributed emissions. For these countries, implicit methodological choices in the input-output model are reflected in both hourly and average yearly emission measures.
%Consumption-based carbon emission measures aim to account for emissions associated with power transmission from distant regions, as opposed to measures which only consider local power generation. In this contribution, we assess an input-output model based on the flow tracing methodology, which connects the location of electricity consumption with the location of generation. Outlining key differences between two different methodological variants of this approach, we report results on consumption-based emission intensities of power generation for European countries from 2016 to 2019. We find that in particular for well connected smaller countries, the consideration of imports has a significant impact on the attributed emissions. For these countries, implicit methodological choices in the input-output model are reflected in both hourly and average yearly emission measures.
\end{abstract}

%\begin{IEEEkeywords}
%cross-border physical flows, principal components analysis, flow tracing
%\end{IEEEkeywords}

\IEEEpeerreviewmaketitle
%
%%%%%-----%%%%%-----%%%%%-----%%%%%-----%%%%%
\section{Introduction}
\label{sec:introduction}
%%%%%-----%%%%%-----%%%%%-----%%%%%-----%%%%%
Decarbonising the power sector plays a key role in Europe’s ambition to be the first climate-neutral continent by 2050~\cite{EC2019}. Monitoring the emissions associated with generation, consumption, and transmission of electricity thus becomes a crucial information that helps steering the measures to implement the envisioned European Green Deal. However, attributing and spatialising emissions in the electricity system is challenged by the fact that generation and consumption are connected via power transmission connections, covering large distances through the electricity grid. Accordingly, generation-based emissions in one country can be very different from the consumption-based emissions in the same country, the latter being determined also by the imports from other countries through the power grid~\cite{Jiusto2006,Wang2017}. Given the increasing integration of European electricity markets, and the rising share of renewable generation located distant from load centers, the corresponding long-distance power flows are expected to be of even larger importance in the future~\cite{tyndp2018}. The incorporation of imports and exports is therefore particularly relevant for the interconnected, heterogeneous European electricity system, with its very different regional generation mixes shaped by local environmental resources, historical trajectories, and current and past policy measures. Recently, this challenge has been addressed using an input-output model adapted to trace power flows through the electricity grid~\cite{electricity_map,DeChalendar2019,Tranberg2019}. It was shown that for both the European and the U.S. power system, the generation-based emissions of the considered regions often substantially differ from the consumption-based emissions. These findings underline the relevance of such flow-based carbon accounting measures, and the need for a deeper understanding of the methodological assumptions and the influence of the underlying data sets. In this contribution, we compare two different variants to couple regional power generation to international transmission flows in the context of the determination of consumption-based carbon emission intensities of European countries. We show that implicit methodological choices can have a significant influence on the resulting characterisation of the system. This emphasises the importance of transparent and well-understood methodologies for the analysis of highly interconnected energy systems, as well as the usage of open data, which together allow a critical assessment and reproduction of key findings for all stakeholders~\cite{Pfenninger2017,Morrison2018}.

This article is structured as follows. After this introduction, Section~\ref{sec:data} reviews the data sources for this study. The subsequent Section~\ref{sec:methods} defines two variants of the flow tracing methodoloy (direct and aggregated coupling), and reviews the concept of consumption-based carbon emission intensities. Results are discussed in Section~\ref{sec:results}, followed by a conclusion in Section~\ref{sec:conclusion}.
%%%%%%%%%%%%%%%%%%%%%%%%%%%%%%%%%
\section{Data}
\label{sec:data}
%%%%%%%%%%%%%%%%%%%%%%%%%%%%%%%%%
%-----%-----%-----%
\begin{figure}[ht]
\centering
\includegraphics[width=0.95\linewidth]{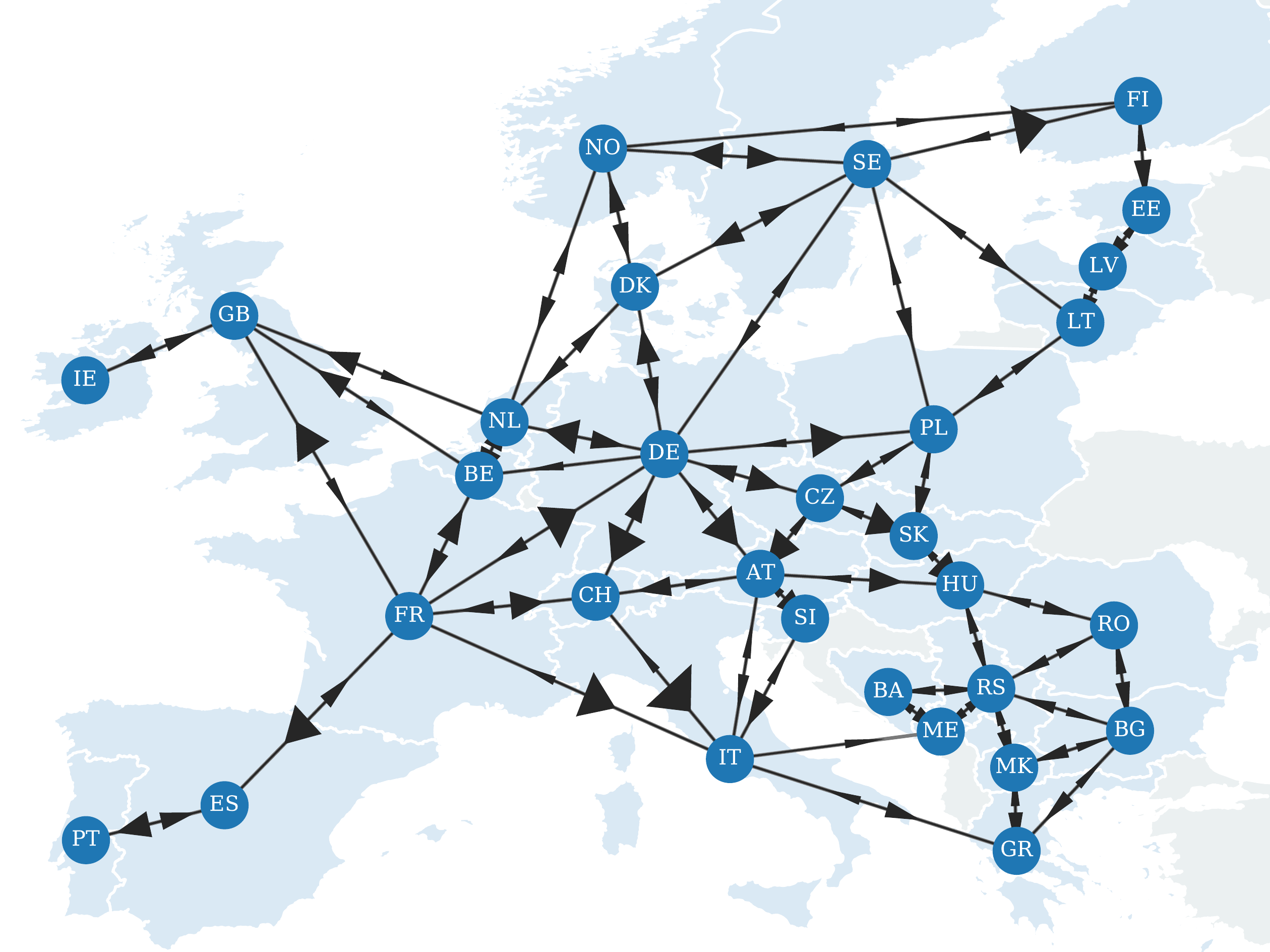}
% where an .eps filename suffix will be assumed under latex, 
% and a .pdf suffix will be assumed for pdflatex; or what has been declared
% via \DeclareGraphicsExtensions.
\caption{Transmission network model with 30~ENTSO-E member countries as nodes, and 57~interconnectors as links. Arrows indicate average hourly cross-border physical flows in 2019 according to data from the ENTSO-E Transparency Page~\cite{transparency}. The largest flows occur from Switzerland to Italy (2.4~GW), from Sweden to Finland (1.8 ~GW), and from France to Germany (1.5~GW).}
\label{fig:topology}
\end{figure}
%-----%-----%-----%
All power system data used in this study is taken from the Transparency Page of the European Network of Transmission System Operators (ENTSO-E)~\cite{transparency}. We use hourly values $g_{(n,\alpha)}(t)$ for generation per production type $\alpha$ in country $n$, total load $d_{n}(t)$, and cross-border physical flows $f_{n\to m}(t)$ for an area comprising 30 out of 35 ENTSO-E member countries (see Fig.~\ref{fig:topology}). The time index $t$ runs over all hours of four years, 2016 - 2019. Production is attributed to nine categories (see Table~\ref{tab:emissions}), with storage dispatch from hydro power classified as generation, and storage charging added to the total load for the specific country. Cross-border physical flows $f_{n\to m}(t)\geq 0$ from country~$n$ to country $m$ are considered for 57~ interconnectors, where the notation allows  positive values only, according to the actual orientation of the flow at time step~$t$. The data is corrected for imbalances, such that conservation of power holds for all countries $n$ (throughout the rest of the article, the time index $t$ is omitted for convenience):
%-----%
\begin{equation}
    \sum_{\alpha}g_{(n,\alpha)}+\sum_{m}f_{m\to n} =
    d_n + \sum_{k}f_{n\to k}~.
\end{equation}
%-----%
%-----%-----%-----%-----%
\begin{table}[ht]
\centering
\begin{tabular}{l d{3.6}} 
 \hline
 Technology & \multicolumn{1}{c}{Intensity [kgCO$_2$eq/MWh]}\\
 \hline
solar	&	0.003599\\
geothermal	&	0.006635\\
wind	&	0.1452\\
nuclear	&	10.37\\
biomass	&	50.47\\
hydro	&	59.49\\
gas	&	530.6\\
oil	&	931.1\\
coal	&	1171\\
% Old values
% solar & 0.00410\\
% geothermal & 0.00664\\
% wind & 0.141\\
% nuclear & 10.3\\
% hydro & 16.2\\
% biomass & 50.9\\
% gas & 583\\
% unknown & 927\\
% oil & 1033\\
% coal & 1167\\
 \hline\\
\end{tabular}
\caption{Average generation-based carbon emission intensities per technology as defined in Eq.~\ref{eq:intensity}. For details consult~\cite{Tranberg2019}.}
\label{tab:emissions}
\end{table}
%-----%-----%-----%-----%
It should be noted that monthly or yearly values for generation, load or cross-border flows published by ENTSO-E~(for instance in~\cite{fact_sheet_2018}) or in other reports~(for instance in~\cite{power_sector_2019}) might differ from aggregated hourly values due to ex-post consolidation taking into account other statistical resources~\cite{hirth2018_TP,Schaefer2019}.\\
Carbon emission intensities $e_{(n,\alpha)}$ for generation per production type $\alpha$ in country $n$ are taken from~\cite{Tranberg2019}. These intensities have been derived from the ecoinvent 3.4 database~\cite{ecoinvent}, taking into account all CO$_2$ equivalent operational emissions occurring over the fuel chain as well as direct emissions from combustion. Accordingly, these intensities differ from values derived from reported emissions in the framework of the EU~ETS mechanism, see for instance~\cite{power_sector_2019}. %Production of unknown type is assumed to be associated to fossil fuel combustion and is thus approximated by values derived from coal, gas, and oil intensities in the individual countries.
%For details consult~\cite{Tranberg2019} and references therein.
Table~\ref{tab:emissions} displays generation-weighted emission intensities per generation type $\alpha$ averaged across countries:
%-----%-----%-----%-----%
\begin{equation}
\label{eq:intensity}
    \langle e_{\alpha}\rangle=\frac{\sum_n e_{(n,\alpha)}g_{(n,\alpha)}}{\sum_n g_{(n,\alpha)}}~.
\end{equation}
%-----%-----%-----%-----%
Renewable generation from solar and wind power generation, for instance, has very low emission intensities mainly associated with maintenance, whereas the high intensities for coal and oil are based on the emissions from fossil fuel combustion~\cite{Tranberg2019}.
%%%%%%%%%%%%%%%%%%%%%%%%%%%%%%%%%
\section{Methods}
\label{sec:methods}
%%%%%%%%%%%%%%%%%%%%%%%%%%%%%%%%%
The flow tracing methodology, which can be interpreted as a multi-regional input-output model~\cite{Tranberg2019,DeChalendar2019}, is described by the following balance equation:
%------%
\begin{align}
    \delta_{n,m}p_{n}^{\mathrm{in}}+\sum_{k}  q_{k,m}f_{k\to n}=
     q_{n,m}p_{n}^{\mathrm{out}} + \sum_{k}q_{n,m}f_{n\to k}~.
\label{eq:partial_flows}
\end{align}
%-----%
Here we represent the meshed European transmission grid as an aggregated network of countries as nodes, which are connected by interconnectors as links. The value $p_n^{\mathrm{in}}$ then refers to the power flow from country~$n$ \emph{into the network} (network inflow), and $p_n^{\mathrm{out}}$ refers to the flow into the country~$n$ \emph{out of the network} (network outflow). From this equation the variables $q_{n,m}$ can be determined, which describe the fraction of outflow $p_n^{\mathrm{out}}$ at node~$n$ originating from the inflow $p_{m}^{\mathrm{in}}$ at node~$m$~\cite{hoersch2018a}. This formulation incorporates the principle of \emph{proportional sharing}, i.e. the equal distribution of the inflow (both the network inflow and the incoming power flow over links) to the outflow (both the network outflow and the outgoing power flow over links)~\cite{Bialek1996,Kirschen1997}. The defining equation~(\ref{eq:partial_flows}) still leaves some degrees of freedom open regarding the coupling to quantities defined \emph{inside} the countries. In the following we define two types of coupling, which determine the nodal inflow and outflow. This in turn affects the definition of the resulting composition of the demand $d_{n,(m,\alpha)}$, which represents the share of the demand at node $n$, that is associated with generation $g_{(n,\alpha)}$ of type $\alpha$ in country $m$.

\paragraph{Direct coupling} This coupling scheme assumes that all generation and demand capacities are directly coupled to the higher-level network. Consequently, a country $n$ with non-zero cross-border physical flows $f_{m\to n}$ and $f_{n\to k}$ has non-zero network inflow (generation, storage discharging) \emph{and} non-zero network outflow (demand, storage charging):
%-----%
\begin{align}
    p_n^{\mathrm{in}} &= g_n = \sum_{\alpha}g_{(n,\alpha)}~,\\
   p_n^{\mathrm{out}} &= d_n~.
\end{align}
%-----%
Here $g_{n}$ denotes the total generation $g_{n}=\sum_{\alpha}g_{(n,\alpha)}$ in country $n$. The share $r_{(n,\alpha)}$ of the network inflow which is associated with generation $g_{(n,\alpha)}$ of generation type $\alpha$ from country $n$ is defined as
%-----%
\begin{align}
   r_{(n,\alpha)}=\frac{g_{(n,\alpha)}}{g_{n}}~.
\end{align}
%-----%
The associated share of demand $d_{m,(n,\alpha)}$ in country $m$ is calculated as
%-----%
\begin{align}
    d_{m,(n,\alpha)}=d_{m}q_{m,n}r_{(n,\alpha)}~.
\end{align}
%-----%
Note that in this coupling scheme, the tracing methodology is applied to the entire generation mix, since all generators are assumed to be directly coupled to the  network. The composition of the demand in a country is thus determined by the generation mix in all countries upstream in the flow pattern in the overall system. This type of coupling has been used for instance  in~\cite{DeChalendar2019,Tranberg2019}.
%%%
\paragraph{Aggregated coupling} This coupling scheme assumes that all generation and demand is aggregated inside the node, and only the net import $IM_{n}$ or net export $EXP_{n}$ is connected to the higher-level network. Consequently, a node will \emph{either} be associated with network inflow (net exporting country) \emph{or} network outflow (net importing country):
%-----%
\begin{align}
    p_n^{\mathrm{in}} &= EXP_{n} = \max\left(g_n-d_n,0\right)~,\\
   p_n^{\mathrm{out}} &= IM_{n} = \max\left(d_n-g_n,0\right)~.
\end{align}
%-----%
The share $r_{(n,\alpha)}$ of the network inflow which is associated with generation $g_{(n,\alpha)}$ of type $\alpha$ in country $n$ is again given by
%-----%
\begin{align}
   r_{(n,\alpha)}=\frac{g_{(n,\alpha)}}{g_{n}}~.
\end{align}
%-----%
The associated share of demand $d_{m,(n,\alpha)}$ has to take into consideration that internal generation first serves internal demand:
%-----%
\begin{align}
    d_{m,(n,\alpha)}=\delta_{m,n}r_{(m,\alpha)}\left(d_{m}-p_m^{\mathrm{out}}\right)+q_{m,n}r_{(n,\alpha)}p_{m}^{\mathrm{out}}~.
\end{align}
%-----%
Note that in this scheme, the tracing methodology is applied only to the net import/net export of the individual countries, not to the entire generation as in the direct coupling scheme. It follows, that for net exporting countries the demand is only associated with internal generation, whereas the generation from net importing countries does not affect the composition of the demand in downstream countries. This type of coupling for instance has been used in~\cite{Schaefer2019,Tranberg2015}.

The \emph{generation-based} carbon intensity of country~$n$ is defined as follows:
\begin{align}
    e_{n}^{\mathcal{G}}=\frac{\sum_{\alpha}e_{(n,\alpha)}g_{(n,\alpha)}}{\sum_{\alpha}g_{(n,\alpha)}}~.
\end{align}
These intensities only refer to the generation inside country~$n$ and do not take into account any imports or exports. Corresponding values can be calculated for individual time steps~$t$, or as generation-weighted average values over the entire period under consideration. In contrast, \emph{consumption-based} carbon intensities take into account imports and exports through the flow tracing methodology reviewed above~\cite{Tranberg2019}:
\begin{align}
    e_{m}^{\mathcal{C}}=\frac{\sum_{\alpha,n}e_{(n,\alpha)}d_{m,(n,\alpha)}}{d_m}~.
\end{align}
Consumption-based emission intensities based on the direct coupling methodology are visualised by the \emph{electricityMap} project~\cite{electricity_map}.
%%%%%%%%%%%%%%%%%%%%%%%%%%%%%%%%%
\section{Results}
\label{sec:results}
%%%%%%%%%%%%%%%%%%%%%%%%%%%%%%%%%
%-----%-----%-----%-----%
\begin{figure}[ht]
\centering
\includegraphics[width=\linewidth]{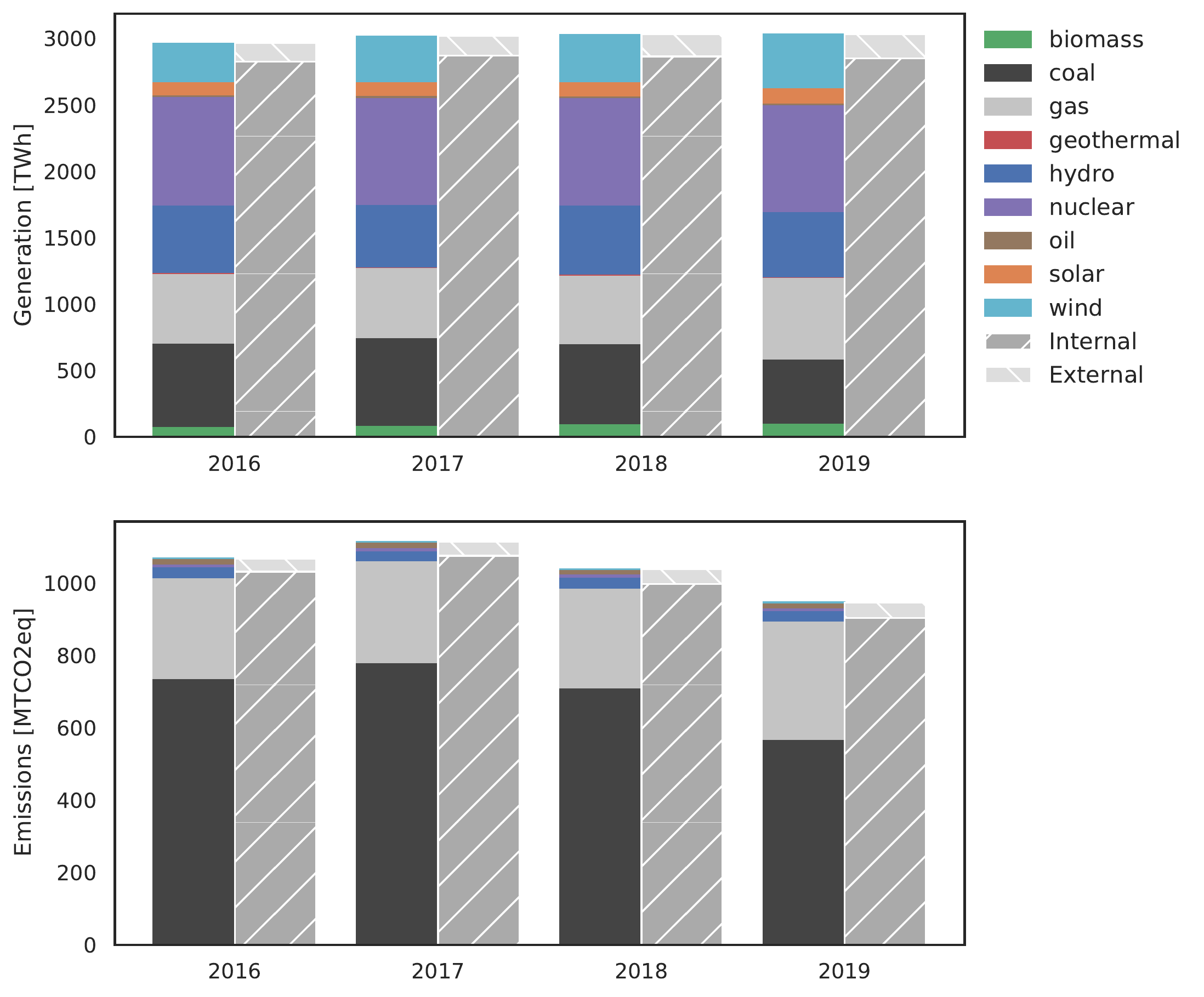}
\caption{Top: Mix of total generation for the years 2016 to 2019 aggregated over 30 ENTSO-E member countries. Bottom: Total generation-based emissions for the years 2016 to 2019. Dark grey values show the aggregates over internal generation (that is, excluding net exports), light grey represents aggregates over external generation (that is, net imports only).}
\label{fig:total}
\end{figure}
%-----%-----%-----%-----%
Figure~\ref{fig:total} shows the evolution of the aggregated yearly generation $\sum_{n}g_{(n,\alpha)}$ and generation-based emissions $\sum_{n}e_{(n,\alpha)}g_{(n,\alpha)}$ of the 30 ENTSO-E member countries considered for this study. Total generation remains almost constant over these four years, whereas the generation mix shows some evolution, notably a decrease in power generation from coal, and an increase in gas and renewable power generation. These changes are reflected in a reduction of about $11\%$ of the generation-based emissions from 2016 to 2019. This increase is mostly due to a decrease of about $23\%$ in emissions associated with power generation from coal, which is only partially offset by an increase of emissions from gas power generation. Note that these values are based on non-consolidated hourly data from the ENTSO-E Transparency Page and operational emission intensities as stated in~\cite{Tranberg2019}. Differences to European power sector carbon dioxide emissions reported in the literature based on consolidated yearly data and combustion-related emissions can be observed, for instance due to missing data for smaller gas power plants or misclassified generation in the hourly data~\cite{power_sector_2019}. Figure~\ref{fig:total} also shows the relation between the internal generation, which is total generation minus total net exports or equivalently total consumption minus total net imports, and external generation, which is total net exports or equivalently total net imports:
%-----%
\begin{align}
    g^{\mathrm{INT}} &=\sum_{n} \left(g_{n}-EXP_{n}\right)=
    \sum_{n}\left(d_{n}-IM_{n}\right)~,\\
    g^{\mathrm{EXT}} &=\sum_{n} EXP_{n}=
    \sum_{n}IM_{n}~.\\
    g^{\mathrm{TOT}} &=\sum_{n}g_{n}=g^{\mathrm{INT}} + g^{\mathrm{EXT}}~.
\end{align}
%-----%
Overall, the share of the total external generation is of the order of $5\%$ to $6\%$, increasing from $4.7\%$ in 2016 to $6.1\%$ in 2019, with the associated emissions increasing from $3.1\%$ to $4.5\%$ of total emissions. It should be emphasised that the aggregated coupling scheme of the flow tracing algorithm only incorporates this external part of the total generation (net imports and net exports), whereas the direct coupling takes the whole generation into account.
%-----%-----%-----%-----%
\begin{figure}[ht]
\centering
\includegraphics[width=\linewidth]{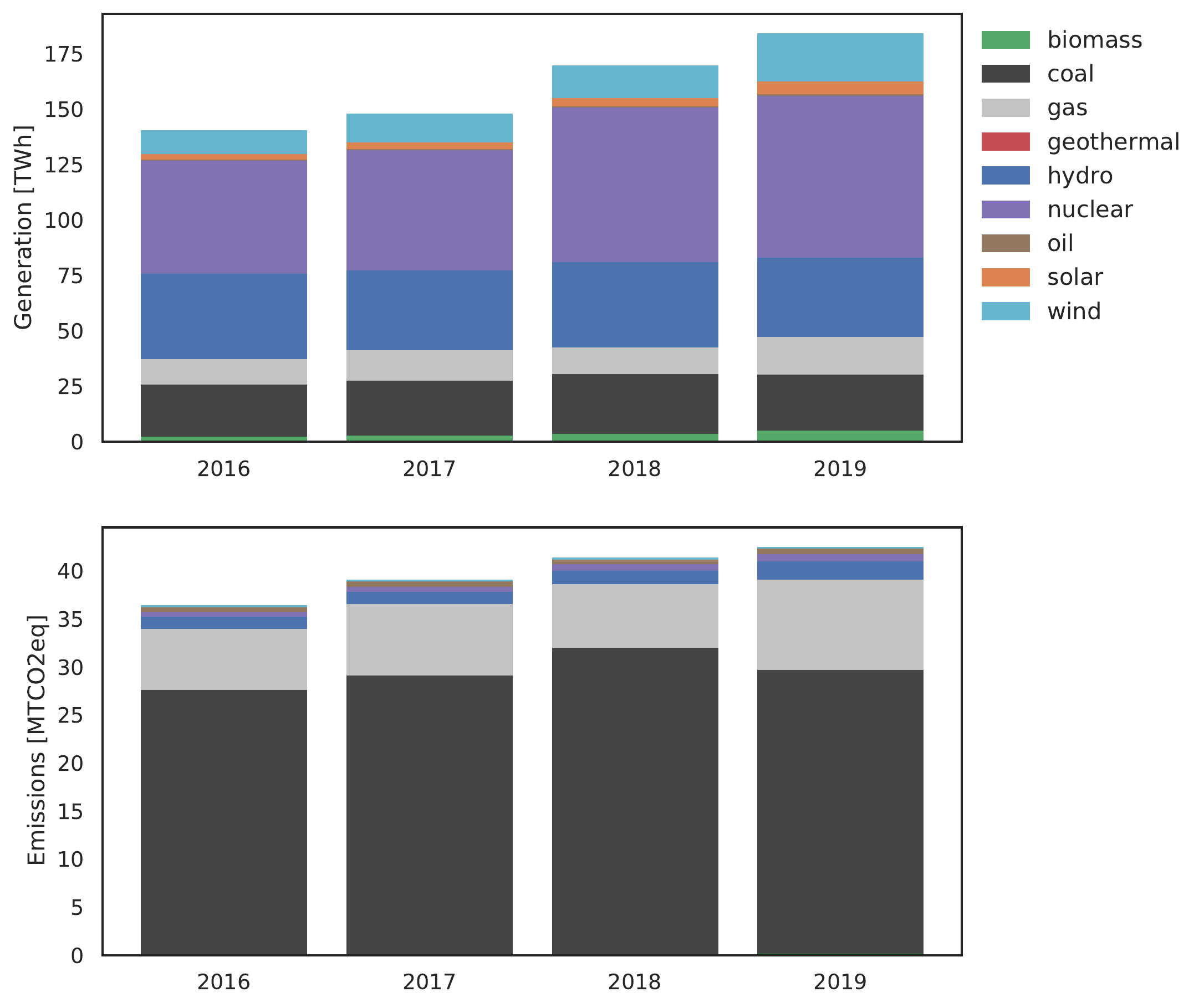}
\caption{Evolution of the total external generation mix (top) and associated emissions (bottom) from 2016 to 2019.}
\label{fig:external}
\end{figure}
%-----%-----%-----%-----%
Figure~\ref{fig:external} displays the total external generation mix and the associated emissions for the period 2016-2019. It can be observed that the increase from $4.7\%$ to $6.1\%$ of total generation corresponds to an absolute increase of about $44$~GWh from 2016 to 2019, corresponding to $32\%$ of the value in 2016. The associated emissions increase by around $6.1$ MtCO$_2$eq in this period, or around $17\%$ of the value in 2016. This indicates that the emission intensity of external generation in general is smaller than the emission intensity of the total generation, which is confirmed by the results shown in Table~\ref{tab:intensities}.
%-----%-----%-----%-----%
\begin{table}[ht]
\centering
\begin{tabular}{l r r r} 
 \hline
  & \multicolumn{3}{c}{Emission intensity}\\
  & \multicolumn{3}{c}{[gCO$_2$eq/kWh]}\\
 Year & Total & Internal & External\\
 \hline
2016 & 359.8 & 364.8 & 259.2 \\
2017 & 369.9 & 374.3 & 264.1 \\
2018 & 342.2 & 348.1 & 243.9 \\
2019 & 311.6 & 316.8 & 230.8 \\
 \hline\\
\end{tabular}
\caption{Average generation-based carbon emission intensities for total, external, and internal generation.}
\label{tab:intensities}
\end{table}
These differences in carbon emission intensities for total, internal, and external generation are due to the different generation mixes. Table~\ref{tab:mix} compares the power mix for the total generation (which is taken into account by the direct coupling scheme) and for the internal generation (which is relevant for the aggregated coupling scheme) for 2019.
%-----%-----%-----%-----%
\begin{table}[ht]
\centering
\begin{tabular}{l r r} 
 \hline
  & \multicolumn{2}{c}{Generation mix}\\
 Technology & Total & External\\
 \hline
solar & 3.8\% & 3.1\%\\
wind & 13.1\% & 11.4\%\\
nuclear & 26.6\% & 39.7\%\\
hydro & 16.2\% & 19.5\% \\
gas & 20.3\% & 9.3\%\\
coal & 15.9\% & 13.7\%\\
other & 4.2\% & 3.3\%\\
 \hline\\
\end{tabular}
\caption{Power mix of total and external generation in 2019. Biomass, geothermal, and oil has been aggregated as `other'. Recall that `hydro' also includes hydro storage dispatch.}
\label{tab:mix}
\end{table}
%-----%-----%-----%-----%
The most significant differences are the higher share of nuclear power in the external generation mix (external:~$39.7\%$, total:~$26.6\%$), to a large degree related to the nuclear power generation of the net exporting country France, and the lower share of gas power (external:~$9.3\%$, total:~$20.3\%$). In combination with the slightly lower share of coal power generation in the external mix (external:~$13.7\%$, total:~$15.9\%$), this leads to the lower carbon emission intensity of external generation compared to total or internal generation (see Table~\ref{tab:intensities}).

\begin{figure}[ht]
\centering
\includegraphics[width=0.475\linewidth]{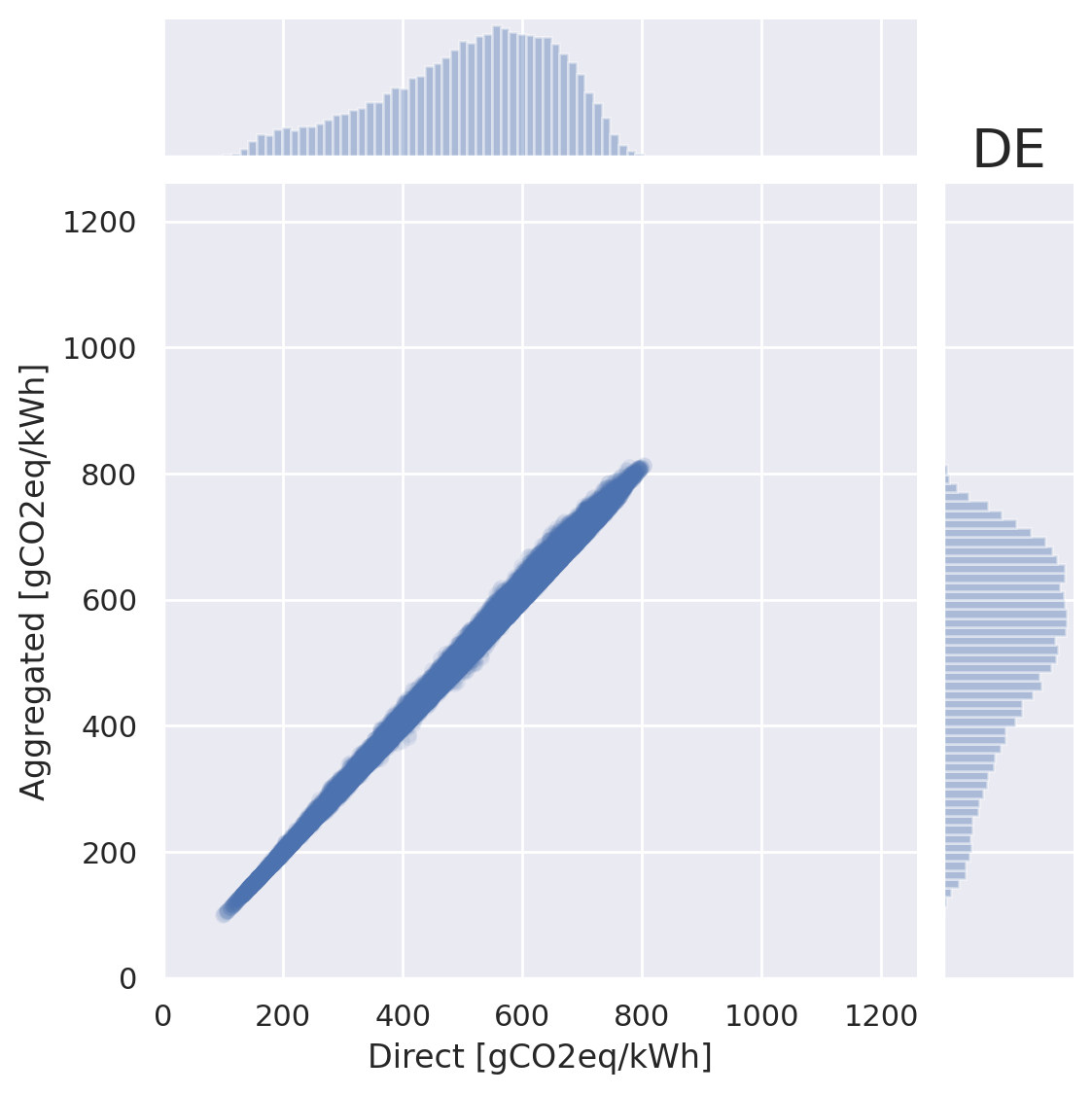}
\includegraphics[width=0.475\linewidth]{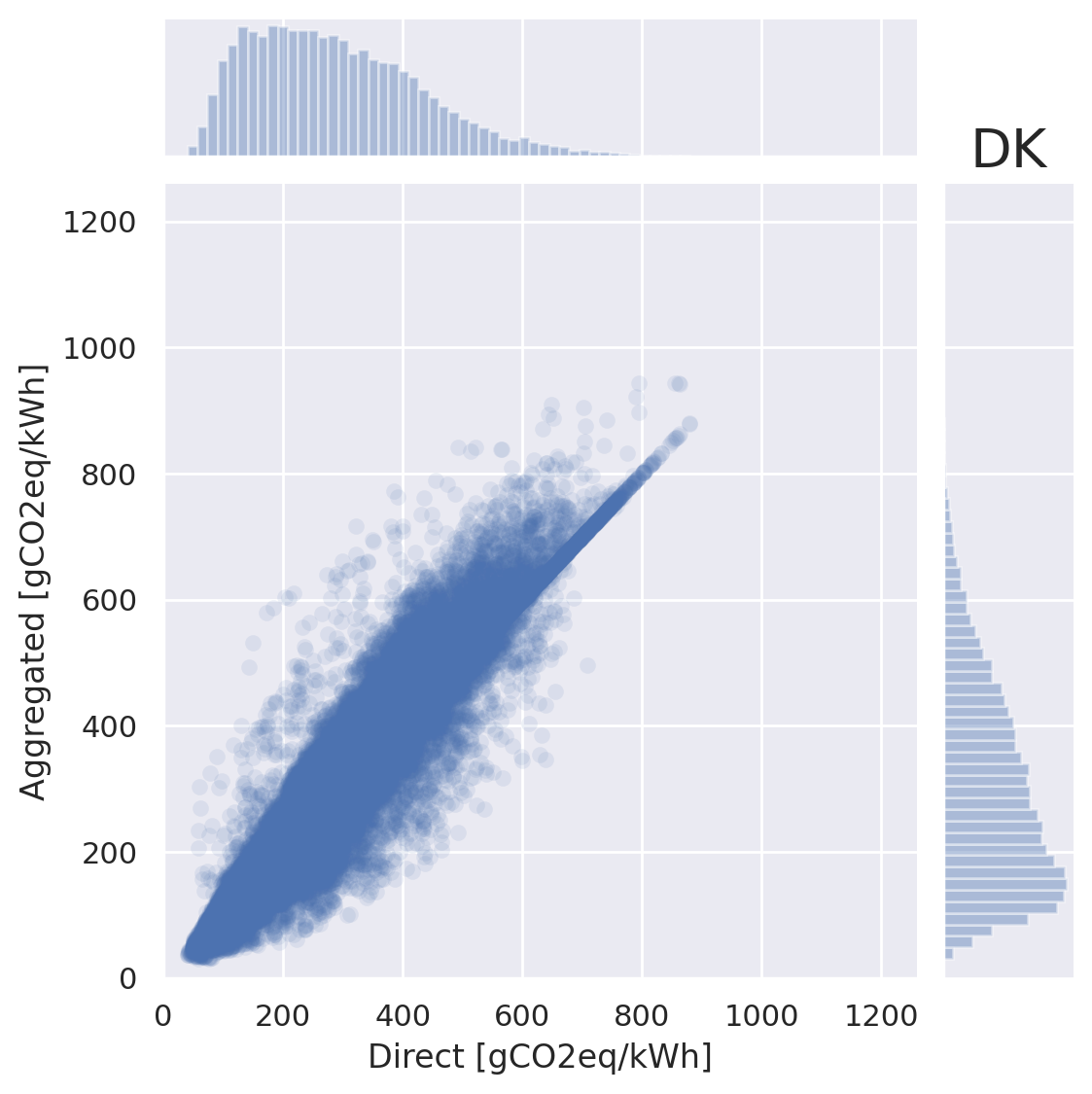}\\
\includegraphics[width=0.475\linewidth]{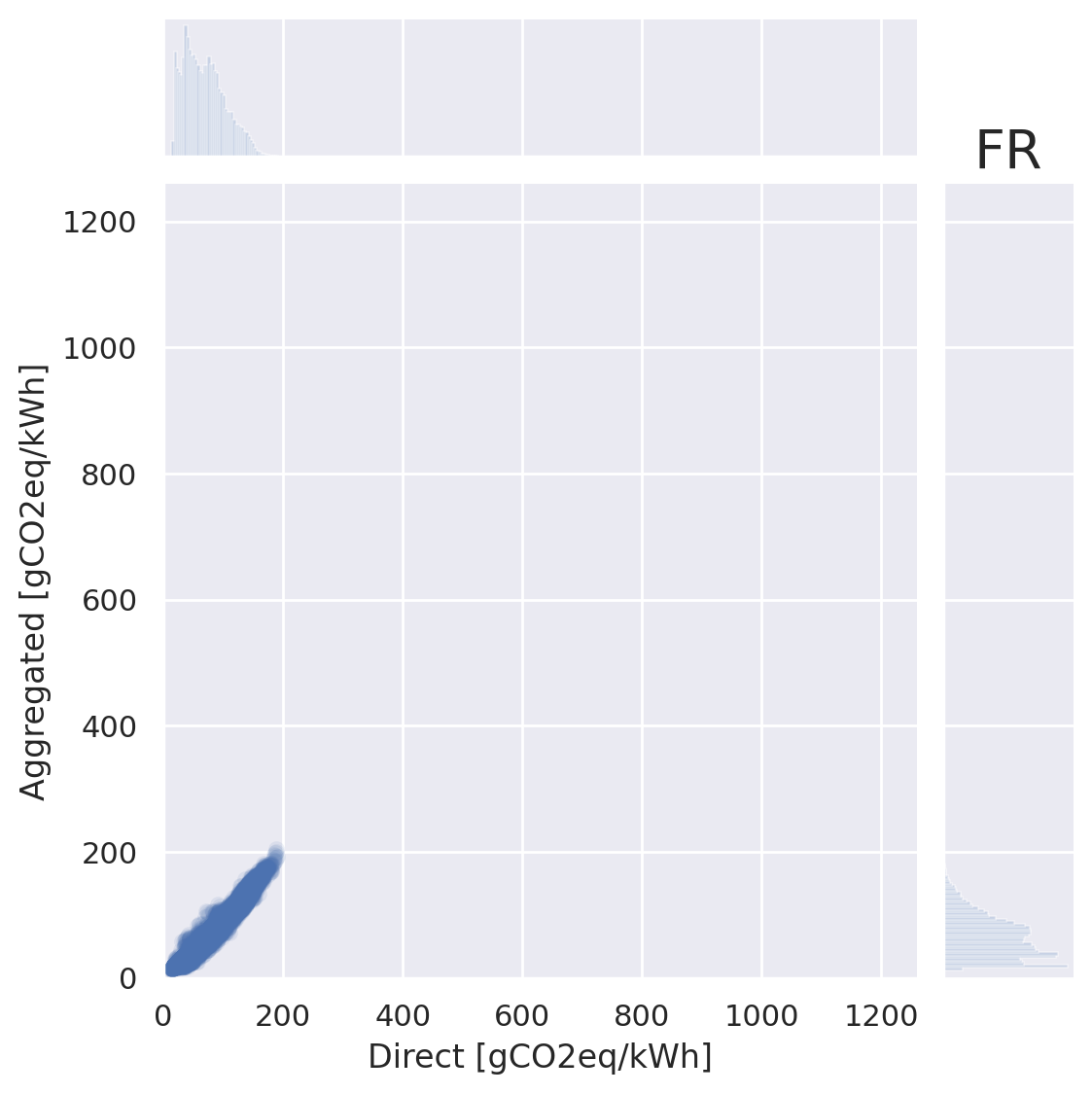}
\includegraphics[width=0.475\linewidth]{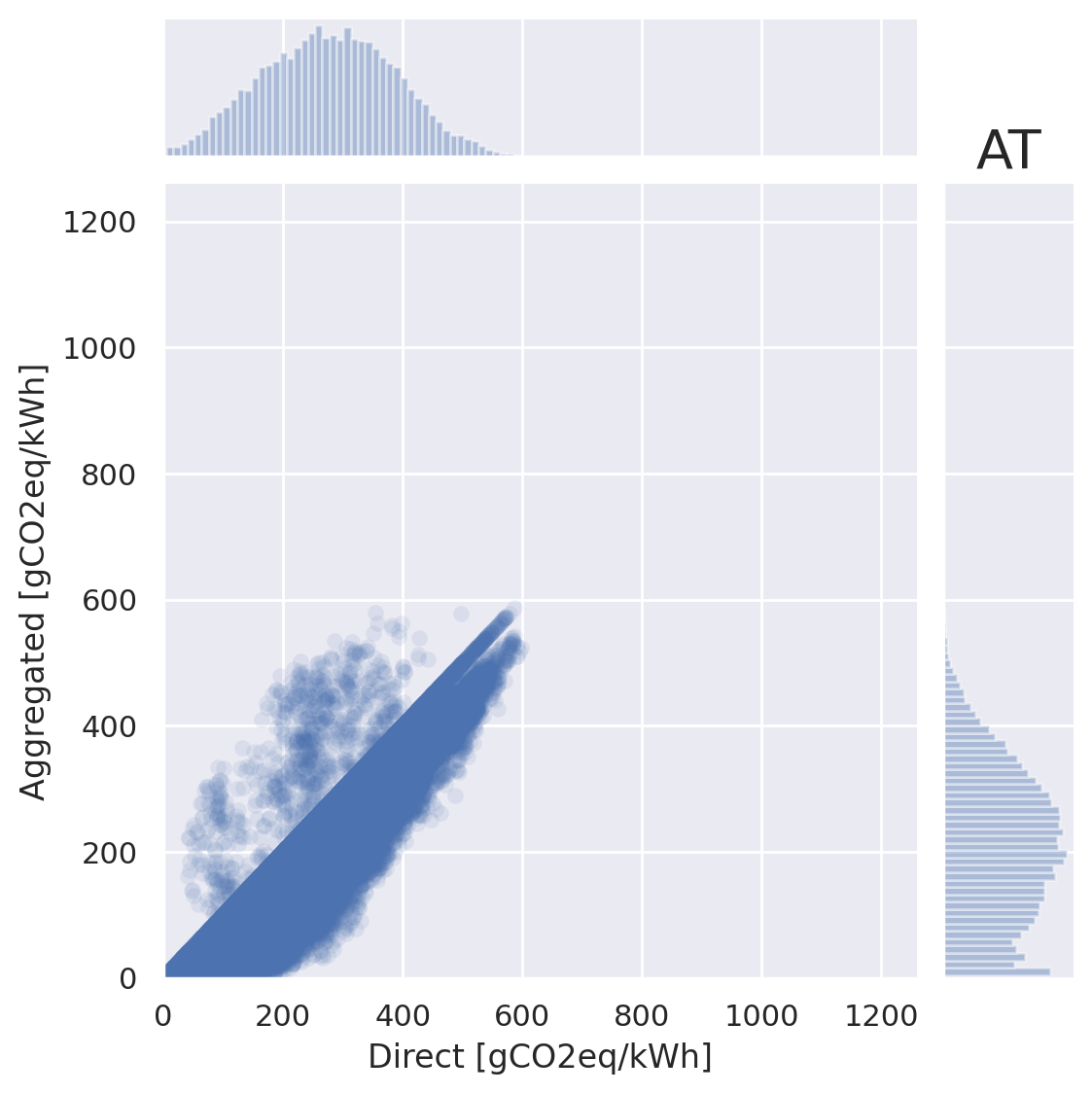}
\caption{Consumption-based emission intensities in gCO$_2$eq/kWh based on the direct (x-axis) and the aggregated (y-axis) coupling scheme. Each dot corresponds to one hour in 2019. The distribution of these intensities over the year is visualised in the margins, respectively. Top left: Germany. Top right: Denmark. Bottom left: France. Bottom right: Austria.}
\label{fig:direct_aggregated}
\end{figure}

\begin{figure}[ht]
\centering
\includegraphics[width=0.95\linewidth]{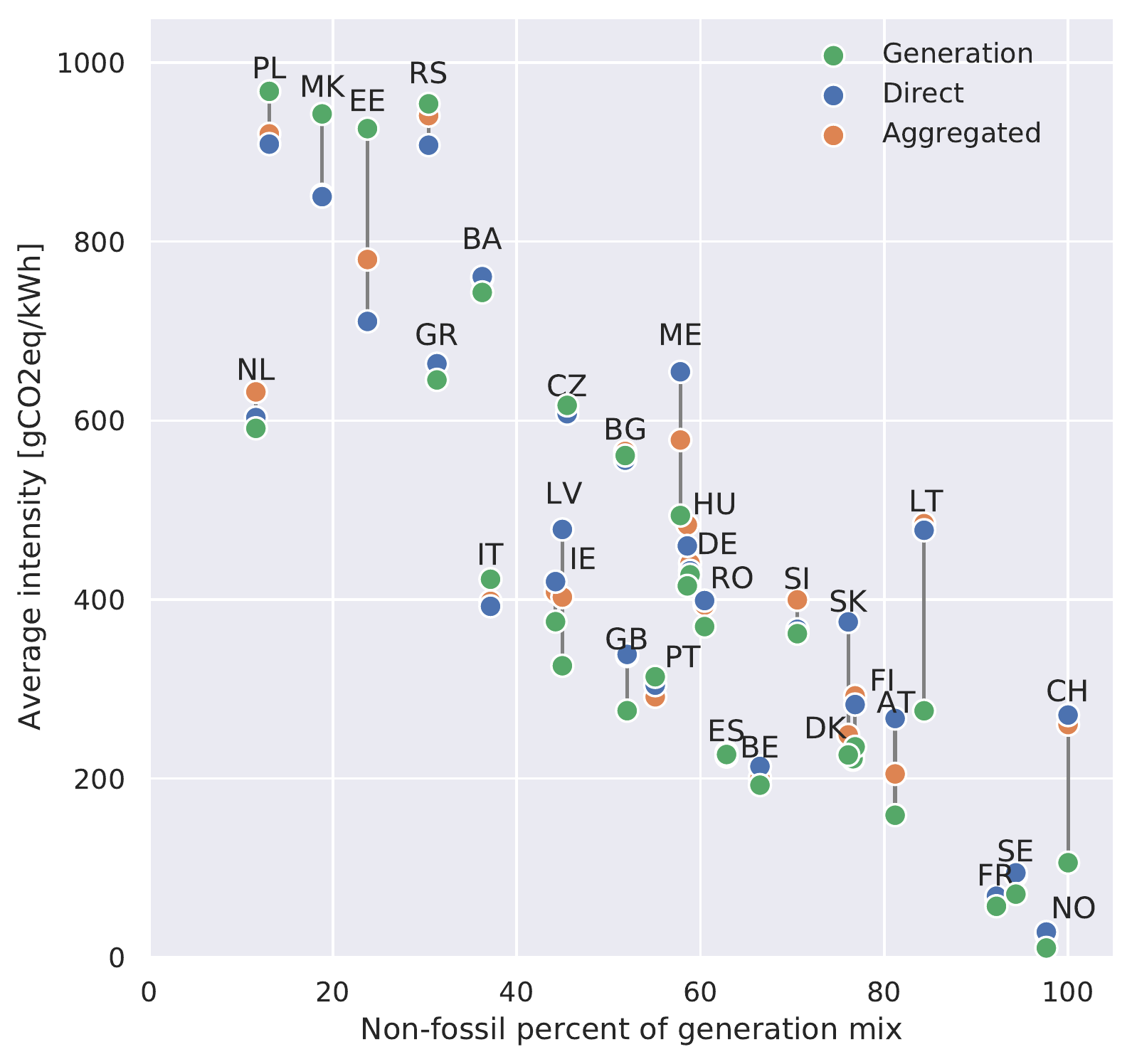}
% where an .eps filename suffix will be assumed under latex, 
% and a .pdf suffix will be assumed for pdflatex; or what has been declared
% via \DeclareGraphicsExtensions.
\caption{Average generation- and consumption-based emission intensities for 30 ENTSO-E member countries in 2019, in comparison to their local share of non-fossil power generation. Green: generation-based. Blue: consumption-based, direct coupling scheme. Orange: generation-based, aggregated coupling scheme. Figure adapted from~\cite{Tranberg2019}.}
\label{fig:scatter}
\end{figure}

The differences in the power mix and the resultant emission intensities for the total and the external generation affect the determination of the consumption-based emission intensities based on the direct and the aggregated coupling scheme, respectively. Figure~\ref{fig:direct_aggregated} shows hourly consumption-based emission intensities for direct vs. aggregated coupling for four selected countries (Germany, France, Denmark, Austria) in 2019. For Germany and France, both schemes lead to very similar results. Representing the countries with the largest generation amongst the ENTSO-E member countries, for these countries local generation is dominant in comparison with imports, which reduces the impact of differences between both coupling schemes. In contrast, for smaller countries with significant cross-border power flows like Denmark or Austria, imports are significant in comparison with local generation. According to~\cite{fact_sheet_2018}, the sum of incoming cross-border physical flows for Germany in 2018 was around 31.5~TWh in comparison to 597.6~TWh of local net generation (France: 13.5~TWh inflow vs. 548.6~TWh net generation), whereas for Denmark the inflows of 15.6~TWh and outflows of 10.4~TWh compare to 28.9~TWh of local net generation (Austria: 29.4~TWh inflows and 19.1~TWh outflows vs. 67.5~TWh net generation). Consequently, for such countries with significant cross-border flows compared to local net generation, the methodological choices for the consideration of emissions associated with imports are relevant. This is visible in Figure~\ref{fig:direct_aggregated}, which shows significant differences in hourly consumption-based emission intensities for Denmark and Austria, based on the direct and the aggregated coupling scheme, respectively.
%-----%-----%-----%-----%
%-----%-----%-----%-----%
Although, in particular for well-connected smaller countries, the hourly values of consumption-based emission intensities thus significantly depend on the choice of the coupling scheme, these results might average out over time. Figure~\ref{fig:scatter} compares generation-based and consumption-based emission intensities (direct and aggregated coupling) for 30 ENTSO-E member countries in 2019 in relation to the share of the non-fossil share of generation in the specific countries (adapted from~\cite{Tranberg2019}). As intuitively expected, countries with a high share of non-fossil generation show comparatively lower emission intensities for the generation-based measures, whereas countries with a high share of fossil generation show lower emissions from consumption-based measures~\cite{Tranberg2019}. It is apparent that also for these average intensities, differences due to the coupling scheme can be observed. Under the aggregated coupling scheme, Austria, for instance, shows an average consumption-based emission intensity which is closer to the generation-based intensity when compared to the value resulting from the direct coupling. This is due to transient flows, which under the direct scheme distribute Austria's low carbon hydro power generation downstream, whereas this generation is accounted predominantly internally to Austria under the aggregated scheme.
%-----%-----%-----%-----%
%-----%-----%-----%-----%
%%%%%-----%%%%%-----%%%%%-----%%%%%-----%%%%%
\section{Conclusion}
\label{sec:conclusion}
%%%%%-----%%%%%-----%%%%%-----%%%%%-----%%%%%
When assessing the carbon footprint of regional electricity consumption, generation-based measures in general do not consider the role of imports and exports. In particular, when net imports or transient flows are significant in comparison with local generation, consumption-based emission measures incorporating power transmissions can lead to very different estimates. In this contribution, we analyse the influence of specific methodological choices in the context of a tracing approach to carbon emission accounting in the European electricity system. One variant, called direct coupling, assumes that a region is immediately embedded into the power grid, and thus all generation and load is directly coupled to the network representation. The opposing perspective is represented by the so-called aggregated coupling, where -- from the network perspective -- a region is assumed to be situated on another level than the higher-level grid. All generation and load, thus, is balanced in the region first, and only remaining net exports or net imports are coupled to the network. For the application to a network of 30 ENTSO-E member countries, we observe that total net imports/exports show a different power mix than total generation, leading to changes in the estimated consumption-based emission intensities under both coupling schemes, in particular for smaller, highly connected countries. These finding highlight the necessity of clearly understanding and communicating the implicit methodological choices inherent in such consumption-based emission measures. A clear suggestion about which of these two perspectives, direct vs. aggregated coupling, should be selected for the analysis of a specific system mostly depends on the grid structure and the spatial scale of the analysis. In this context, further understanding of the influence of the spatial scale in the network representation on the flow tracing results is needed~\cite{Schaefer2017_scaling}.

Our analysis sets the focus on the methodological details of incorporating power transmission in consumption-based emission measures associated with electricity generation. From a broader perspective, establishing such measures is even more dependent on the availability and quality of the underlying data. For instance, although transmission operators already provide detailed power sector data through the ENTSO-E Transparency Platform, this data is still far from perfect, and in general needs different levels of corrections. Improving data quality and establishing common best practices for taking into account still existing inadequacies thus is a necessary complement to methodological studies as presented in this contribution.
%The insights yielded from such and related studies can emphasize how the trade of energy between European countries makes use of the common power grid, which could support public acceptance of infrastructure projects. Furthermore, an understanding of relevant patterns can provide relevant information to market participants, thus promoting the economic efficiency of the energy system.

\ifCLASSOPTIONcaptionsoff
  \newpage
\fi



%\bibliographystyle{IEEEtran}
\bibliographystyle{IEEEtran}
% \nocite{*}
\bibliography{references}

\begin{thebibliography}{10}
\providecommand{\url}[1]{#1}
\csname url@samestyle\endcsname
\providecommand{\newblock}{\relax}
\providecommand{\bibinfo}[2]{#2}
\providecommand{\BIBentrySTDinterwordspacing}{\spaceskip=0pt\relax}
\providecommand{\BIBentryALTinterwordstretchfactor}{4}
\providecommand{\BIBentryALTinterwordspacing}{\spaceskip=\fontdimen2\font plus
\BIBentryALTinterwordstretchfactor\fontdimen3\font minus
  \fontdimen4\font\relax}
\providecommand{\BIBforeignlanguage}[2]{{%
\expandafter\ifx\csname l@#1\endcsname\relax
\typeout{** WARNING: IEEEtran.bst: No hyphenation pattern has been}%
\typeout{** loaded for the language `#1'. Using the pattern for}%
\typeout{** the default language instead.}%
\else
\language=\csname l@#1\endcsname
\fi
#2}}
\providecommand{\BIBdecl}{\relax}
\BIBdecl

\bibitem{EC2019}
{European Commission}, ``{The European Green Deal},'' 2019.

\bibitem{Jiusto2006}
S.~Jiusto, ``{The differences that methods make: Cross-border power flows and
  accounting for carbon emissions from electricity use},'' \emph{Energy
  Policy}, vol.~34, no.~17, pp. 2915--2928, 2006.

\bibitem{Wang2017}
F.~Wang, J.~Shackman, and X.~Liu, ``{Carbon emission flow in the power industry
  and provincial CO2 emissions: Evidence from cross-provincial secondary energy
  trading in China},'' \emph{Journal of Cleaner Production}, vol. 159, pp.
  397--409, 2017.

\bibitem{tyndp2018}
ENTSO-E, ``{TYNDP 2018 Executive Report},'' 2018.

\bibitem{electricity_map}
``{electricityMap},'' \url{https://www.electricitymap.org/}.

\bibitem{DeChalendar2019}
J.~A. de~Chalendar, J.~Taggart, and S.~M. Benson, ``{Tracking emissions in the
  US electricity system},'' \emph{Proceedings of the National Academy of
  Sciences of the United States of America}, vol. 116, no.~51, pp.
  25\,497--25\,502, 2019.

\bibitem{Tranberg2019}
B.~Tranberg, O.~Corradi, B.~Lajoie, T.~Gibon, I.~Staffell, and G.~B. Andresen,
  ``Real-time carbon accounting method for the {European} electricity
  markets,'' \emph{Energy Strategy Reviews}, vol.~26, p. 100367, 2019.

\bibitem{Pfenninger2017}
S.~Pfenninger, J.~DeCarolis, L.~Hirth, S.~Quoilin, and I.~Staffell, ``{The
  importance of open data and software: Is energy research lagging behind?}''
  \emph{Energy Policy}, vol. 101, pp. 211--215, 2017.

\bibitem{Morrison2018}
R.~Morrison, ``{Energy system modeling: Public transparency, scientific
  reproducibility, and open development},'' \emph{{Energy Strategy Reviews}},
  vol.~20, pp. 49--63, 2018.

\bibitem{transparency}
``{ENTSO-E Transparency Platform},'' \url{https://transparency.entsoe.eu}.

\bibitem{fact_sheet_2018}
ENTSO-E, ``{Statistical Factsheet 2018},'' 2019.

\bibitem{power_sector_2019}
{Agora Energiewende} and Sandbag, ``{The European Power Sector in 2019:
  Up-to-Date Analysis on the Electricity Transition},'' 2020.

\bibitem{hirth2018_TP}
L.~Hirth, J.~M{\"{u}}hlenpfordt, and M.~Bulkeley, ``{The ENTSO-E Transparency
  Platform – A review of Europe's most ambitious electricity data
  platform},'' \emph{Applied Energy}, vol. 225, pp. 1054--1067, 2018.

\bibitem{Schaefer2019}
M.~Sch{\"{a}}fer, F.~Hofmann, H.~Abdel-Khalek, and A.~Weidlich, ``{Principal
  Cross-Border Flow Patterns in the European Electricity Markets},'' in
  \emph{International Conference on the European Energy Market, EEM2019}, 2019.

\bibitem{ecoinvent}
G.~Wernet, C.~Bauer, B.~Steubing, E.~Reinhard, E.~Moreno-Ruiz, and B.~Weidema,
  ``{The ecoinvent database version 3 (part I): overview and methodology},''
  \emph{International Journal of Life Cycle Assessment}, vol.~21, no.~9, pp.
  1218 -- 1230, 2016.

\bibitem{hoersch2018a}
J.~H{\"{o}}rsch, M.~Sch{\"{a}}fer, S.~Becker, S.~Schramm, and M.~Greiner,
  ``{Flow tracing as a tool set for the analysis of networked large-scale
  renewable electricity systems},'' \emph{International Journal of Electrical
  Power {\&} Energy Systems}, vol.~96, pp. 390--397, 2018.

\bibitem{Bialek1996}
J.~Bialek, ``{Tracing the flow of electricity},'' \emph{IEE Proceedings -
  Generation, Transmission and Distribution}, vol. 143, no.~4, p. 313, 1996.

\bibitem{Kirschen1997}
D.~Kirschen, R.~Allan, and G.~Strbac, ``{Contributions of individual generators
  to loads and flows},'' \emph{IEEE Transactions on Power Systems}, vol.~12,
  no.~1, pp. 52--60, 1997.

\bibitem{Tranberg2015}
B.~Tranberg, A.~B. Thomsen, R.~A. Rodriguez, G.~B. Andresen, M.~Sch{\"{a}}fer,
  and M.~Greiner, ``{Power flow tracing in a simplified highly renewable
  European electricity network},'' \emph{New Journal of Physics}, vol.~17,
  no.~10, 2015.

\bibitem{Schaefer2017_scaling}
M.~Sch{\"{a}}fer, S.~{Bugge Siggaard}, K.~Zhu, C.~{Risager Poulsen}, and
  M.~Greiner, ``{Scaling of transmission capacities in coarse-grained renewable
  electricity networks},'' \emph{EPL}, vol. 119, no.~3, 2017.

\end{thebibliography}

\end{document}